\documentclass[prx,twocolumn,nofootinbib,longbibliography,preprintnumbers,amsmath,amssymb,10pt]{revtex4}
\UseRawInputEncoding
\usepackage{bm,graphicx,multirow,subfigure,cases}

\usepackage[shortlabels]{enumitem}

\usepackage{braket}

\def\source#1#2#3#4{{\it #1}~{\bf #2}, #3 (#4)}
\def\be{\begin{equation}}
\def\ee{\end{equation}}

\def\ie{i.e.}
\def\CH{\cal H}
\def\kpsi{\ket{\psi}}

\def\alk{\alpha_k}

\def\ak{a_k}

\def\braket#1#2{\langle #1 | #2 \rangle}

\begin{document}


\title{The Born Rule - Axiom or Result?}
\author{Jay Lawrence}
	\email{jay.lawrence.@dartmouth.edu}
	 \affiliation{Department of Physics and Astronomy, Dartmouth College, Hanover, NH, USA} 
	  \affiliation{The James Franck Institute. The University of Chicago, Chicago, IL, USA}  
\author{Philip Goyal}
    \email{pgoyal@albany.edu}
    \homepage[Website: ]{https://www.philipgoyal.org}
    \affiliation{Department of Physics, University at Albany~(SUNY), NY, USA}
\date{\today}
\begin{abstract}
 
\noindent 
The Born rule is part of the collapse axiom in the standard version of quantum 
theory, as presented by standard textbooks on the subject.   We show here that
its signature quadratic dependence follows from a single additional physical 
assumption beyond the other axioms - namely, that the probability of a particular 
measurement outcome (the state $\phi_k$, say) is independent of the choice of 
observable to be measured, so long as one of its eigenstates corresponds to 
that outcome.  We call this assumption ``observable independence.''  As a 
consequence, the Born rule cannot be completely eliminated from the list of 
axioms, but it can, in principle, be reduced to a more physical statement.  Our 
presentation is suitable for advanced undergraduates or graduate students 
who have taken a standard course in quantum theory.  It does not depend on 
any particular interpretation of the theory.  
  
\end{abstract}

\maketitle

\section{Introduction}

There is a wide range of attitudes regarding the Born rule and its status 
among the axioms of quantum theory.  Some accept it as an axiom, while
others think that it should be (or has been) derived from 
the other axioms.  Still others believe that the truth lies somewhere in 
between - that a derivation is possible, but it requires one or more 
additional assumptions beyond the other axioms.  To better describe these 
positions, let us state the Born rule as it might appear in the context of
the other axioms.

Here are the axioms as they are stated, more or less consistently, by textbooks 
in use over the last seventy years \cite{authors}.  The following represents a 
sort of consensus among those listed, following none in particular.
\medskip

\noindent I.  {\it An isolated system is associated with a Hilbert space $\CH$.   
Its pure states are described by unit vectors $\kpsi$ in $\CH$} (normalized to
$\langle \psi | \psi \rangle = 1$) \cite{ray}.
\medskip

\noindent II.  {\it The evolution of an isolated system is described by a unitary 
transformation:  $\ket{\psi(t)} = U \ket{\psi(0)}$, where $U$ represents the 
solution of the time dependent Schr\"{o}dinger equation, 
$i \hbar \frac{\partial}{\partial t} \kpsi = H \kpsi$, and $H$ is the Hamiltonian 
of the system}. 
\medskip

\noindent If $H$ is time independent, then $U = \exp(-iHt/\hbar)$.  
\medskip

\noindent III.  {\it Observable quantities} $\Omega$ {\it are represented by 
Hermitian  (\ie, linear and self-adjoint) operators on $\CH$.}  
\medskip

\noindent It will be useful to expand a state $\kpsi$ in terms of eigenstates 
of $\Omega$, that is,
\be 
   \kpsi = \sum_i \alpha_i \ket{\phi_i},  \hskip.5truecm \hbox{where}  \hskip.4truecm
   \Omega \ket{\phi_i} = \omega_i \ket{\phi_i}.
\label{expansion}
\ee
The eigenvalues $\omega_i$ are real because $\Omega$ is Hermitian.
\smallskip

\noindent IV.  {\it A measurement of $\Omega$ yields a particular eigenvalue} 
($\omega_k$) {\it with probability}
\be
    p_k = |\braket{\phi_k}{\psi}|^2.  
\label{Born rule}
\ee
This probability formula is called the Born rule - typically not a separate 
axiom, but part of this so-called ``collapse'' axiom. 
\medskip

\noindent V.  {\it The state of the system immediately following the measurement is the
eigenstate $\ket{\phi_k}$ corresponding to the measured outcome $\omega_k$.  A 
subsequent measurement of $\Omega$ reproduces the same outcome with certainty.}
\medskip

\noindent This fifth axiom is called ``the post-measurement state update rule.''  It is 
implicit in the von Neumann measurement formalism \cite{von Neumann}, although 
it is valid only if the measurement is {\it ideal} \cite{ideal}; that is, minimally disturbing 
and repeatable on the same system.  As a common counterexample, measurement 
sometimes destroys the system, as in a photon polarization measurement.  

Axioms I - III are the ``dynamical'' axioms, which describe the continuous, deterministic
evolution of the unobserved system (what von Neumann calls ``type II'' processes); 
while IV and V (``collapse and ``state update'') describe the abrupt and random change 
produced by measurement (which von Neumann calls ``type I'').  The word collapse, as 
used here, refers to the observed facts described by IV and V, and not to any 
interpretation-dependent concept of what might lie behind it.  The Born rule, while part 
of the collapse axiom, bridges the gap separating the two realms of behavior, referring 
only to input and output states, with no description of the processes connecting them.
The fifth axiom is sometimes absorbed into the fourth by saying that the outcome of a
measurement is the output {\it state}, $\ket{\phi_k}$, instead of the eigenvalue 
$\omega_k$.  Sakurai \cite{authors} puts this point most graphically:  ``... the system is 
thrown into an eigenstate $\ket{\phi_k}$ of $\Omega$''.


One cannot dispute the {\it facts} represented by the axioms above, but interpretations 
differ regarding the status of the collapse axioms and the unanswered questions that 
lie behind them.  The Copenhagen Interpretation (and more modern descendants) 
accept collapse as an independent axiom, implying that its underpinnings are not a 
subject for study within quantum theory.  Other interpretations, including 
but not limited to many worlds \cite{Everett}, attempt to derive the collapse 
{\it phenomenon}, as well as the Born rule itself, from the dynamical axioms.  
Decoherence theory, which is not tied to any particular interpretation, provides a formal 
background that describes the collapse phenomenon using axioms I - III (and in this 
way it supports the many worlds interpretation).  However, it does not provide a 
derivation of the Born rule, because the Born rule is assumed implicitly in its formalism.  
\cite{Weinberg} 

Standard quantum mechanics, as practiced by most physicists, accommodates a 
broad range of interpretations between the two extremes above, and it makes use 
of decoherence theory as a practical tool.  In this paper, by accepting the collapse
phenomenon as a fact and focusing just on the Born rule itself, we shall present a 
derivation which is independent of interpretations, and which illustrates the apparent 
necessity of an additional assumption.  Stylistically, for the sake of transparency, we 
will be concise and not comprehensive.  We will tend to emphasize physics over 
mathematics, citing more detailed mathematical treatments as we progress.  
Our aim is to provide an intuitive physical derivation of the quadratic dependence 
of the Born rule, and thus to clarify its status as part axiom and part derived.


In the next section we show that the quadratic dependence of the Born rule
follows from the ``dynamicalÕÕ axioms'' (I - III) plus the assumption of ``operator
independence,'' which we shall describe in more detail.  In Section III we compare 
our derivation with other approaches to illustrate the variety of avenues being 
explored.  We conclude in Sec. IV by reviewing the status of the Born rule as 
we see it.

 
\section{Physical Origin of the Born rule}

The Born rule is defined by Eq. \ref{Born rule}.   The signature quadratic dependence 
is explicit, and considered by many to be synonymous with the Born rule itself.  But we 
draw attention to two other properties which are implicit:   (i) the probability $p_k$ is 
independent of the other (unobserved) expansion coefficients ($\alpha_j$ with 
$j \neq k$), and perhaps less obviously (ii) $p_k$ is independent of the choice of the 
observable ($\Omega$), so long as $\Omega$ has the output state $\ket{\phi_k}$ as 
an eigenstate.   In what follows, we shall first derive the quadratic dependence by 
assuming (i) explicitly.  We shall then assume (ii) explicitly, and derive both (i) and 
the quadratic dependence from it.  So (ii) is the deeper, as well as the more 
physical assumption.  

To set the stage, we begin by making neither assumption - we imagine that the 
probability $p_k$ could, in principle, depend on all the expansion coefficients 
in Eq. \ref{expansion}, with additional dependence on the observable $\Omega$ 
being measured; that is,
\be
   p_k = f_k(\{\alpha_i\},\Omega), \hskip1.5truecm   i = 1,...,d.
\label{general function 1}
\ee
The expansion coefficients refer to the basis of eigenstates of $\Omega$, which 
is called the {\it measurement basis}.

Both of our proofs are based on the invariance of $p_k$ under unitary 
transformations of the complex vector $\{\alpha_i\}$ of expansion coefficients.  
These transformations correspond to different representations of the same 
prepared initial state $\ket{\psi}$ (that is, to different measurement bases).  

As a trivial preliminary example, note that the {\it phases} of the above basis
states [\ie, $\gamma_k$ in $\alpha_k \equiv a_k \exp(\gamma_k)$] can be 
chosen arbitrarily, which means that $p_k$ can only depend on the 
magnitudes $a_k$ :
\be
   p_k = f_k(\{a_i\},\Omega), \hskip1.5truecm   i = 1,...,d,
\label{general function 2}
\ee
and the normalization condition for state $\ket{\psi}$ simplifies (slightly) to
\be
       \sum_{k=1}^d |\alk|^2 =  \sum_{k=1}^d \ak^2 = 1.
\label{norm state}
\ee
But this simplification has a convenient geometrical interpretation:  The unitary
transformation on the vector of complex expansion coefficients, 
\be
    (\alpha_1, .., \alpha_d)^T  \rightarrow  (\alpha_1', ...,\alpha_d')^T,
\label{complex vector}
\ee
reduces to an orthogonal transformation of the vector of moduli; that is,
\be
   {\bf a} = (a_1, ..., a_d)^T \rightarrow  (a_1', .., a_d')^T = {\bf a'}.
\label{real vector}
\ee
This is a rotation on the {\it unit orthant} (the $2^{-d}$ segment of the unit sphere 
in $d$ real dimensions, in which all components are non-negative). 

We now implement assumption (i) by assuming that $p_k$ depends on only the 
single amplitude ($a_k$) corresponding to the outcome state:  $p_k = f(a_k)$.  
We can then write the normalization sum {\it for probabilities} (as distinct from 
that of the {\it state} itself), as
\be
   {\cal N}(\{a_i\}) = \sum_{i=1}^d f(a_i).
\label{normsum}
\ee
Now ${\cal N}$ must be stationary on the unit orthant.  Invoking this condition 
through a Lagrange multiplier $\lambda$, we find that
\be
    {\partial \over \partial a_j} \bigg[  {\cal N}(\{a_i\}) - \lambda \big(\sum_{i=1}^d a_i^2 - 1\big)
    \bigg] = {\partial f \over \partial a_j} - 2 \lambda a_j = 0.
\label{variation1}
\ee
The solution is
\be
   f(x) = \lambda x^2 + \mu,
\label{solution1}
\ee
and this, with the boundary conditions $f(0)=0$ and $f(1)=1$, reduces to the Born rule, 
$f(x) = x^2$.  This is intuitive - it shows that there is no function of a single variable, 
except for $x^2$, for which the probabilities can be normalized together with the state.  

However, there is no a priori justification for the assumption (i) by itself.
It does not follow from the ``dynamical axioms,'' and it begs the question of a physical 
origin.   Assumption (ii) {\it provides} a physical origin - namely, that $p_k$ is 
independent of the choice of observable to be measured, so long as it has the outcome 
state $\phi_k$ as one of its eigenstates.  To implement this assumption mathematically, 
we demand that $p_k$ be invariant under unitary transformations $U_k$ that preserve 
the single modulus $a_k$.  Such transformations produce arbitrary rotations of ${\bf a}$ 
at fixed $a_k$:
\be
   \big( a_k,~\{a_i,~(i\neq k)\} \big) \rightarrow \big( a_k,~\{a_i',~(i \neq k)\} \big),
\label{rotations1}
\ee
or more concisely, introducing the $(d-1)$-tuples $\tilde{\bf a}$ and $\tilde{\bf a}'$ 
perpendicular to the $k$ direction,
\be
   \big( a_k,~\tilde{\bf a} \big)  \rightarrow   \big( a_k,~\tilde{\bf a}' \big).
\label{rotations2}
\ee
Thus $U_k$ rotates $\tilde{\bf a}$ into $\tilde{\bf a}'$ on the orthant of radius 
$\rho = \sqrt{1-a_k^2}$ in $d-1$ dimensions.  The stationarity of $p_k$ subject to this 
constraint is expressed by 
\be
  {\partial \over \partial a_j} \bigg[ p_k - \lambda \big(\sum_{i \neq k} a_i^2 - \rho^2 \big) 
  \bigg] =  \partial p_k / \partial a_j - 2 \lambda a_j =  0,  
\label{variation2}
\ee
for all $j \neq k$.  This has the solution
\be
   p_k = \lambda \sum_{j \neq k} a_j^2  + \mu = \lambda (1- a_k^2) + \mu.
\label{solution2}
\ee
Since this formula holds for any initial choice of $a_k$, it provides the functional form 
of $p_k$.  Applying the boundary conditions as above, we determine the parameters 
$\mu = -\lambda = 1$, resulting in $p_k = a_k^2$.  Thus we have derived the quadratic
dependence on the single expansion coefficient corresponding to the output state.

We should comment that this proof, unlike some others, applies to the case $d  = 2$, 
even though the sum over $i$ (with $i \neq k$) contains only a single term.  

The physical assumption employed above is what we have called ``observable 
independence.''   Essentially the same
assumption made elsewhere has been called non-contextuality, a term which
we have avoided up to now because it has multiple meanings (as spelled out in 
a recent  compilation by Hofer-Szabo \cite{H-S.22}).   Its meaning here refers to
the context of alternate possible outcomes:  $p_k$ does not depend on the
particular identities of these outcomes.   But we  prefer to emphasize the more 
physical underpinning of the Born rule by referring to the measured observable.  
This suggests that the role of $\Omega$ is to generate a sort of prism which 
separates the component eigenstates (for example, into a set of non-overlapping 
paths) prior to detection.  One detects a particular path ($k$); the alternate 
paths are immaterial, as are the alternate eigenvalues of $\Omega$.

\subsection{An Example}

An instructive example is a $J=1$ atom, with eigenstates of $J_z$ written as
$\ket{m}$, with $m = 1$, 0, and $-1.$  But consider the alternative basis set 
consisting of $\ket{\pm} = (\ket{1} \pm \ket{-1})/\sqrt{2}$, and again $\ket{0}$.
The latter are eigenstates of the operator $J_x^2 - J_y^2$.  Whatever the 
initial state $\ket{\psi}$, measurements of the two operators will yield the 
same probability of output for the state $\ket{0}$.  This is what we mean 
by observable independence.

\section{Related Work and Alternative Approaches}

A. M. Gleason is recognized as the first to prove the Born rule \cite{Gleason.57}, 
and his proof utilizes the  other axioms, 
with the added assumption of non-contextuality.  Gleason's argument is couched 
in mathematical terms, and is generally regarded by physicists as difficult to 
interpret.  Many papers have been written since, generalizing and reformulating 
the original proof.  For example, Caves et. al. \cite{Caves} obtained a simpler 
proof by enlarging the class of measurements to include POVMs (positive 
operator-valued measures \cite{NC}), as opposed to just the orthogonal 
projector-valued measures assumed by Gleason.  For a recent approach that 
makes instructive connections between Gleason's results and later formulations, 
we recommend the recent article by Wright and Weigert \cite{Wright-Weigert}, 
and its references.  We also recommend an approachable article by Lugiurato 
and Smerzi \cite{LS.12}, which introduces the more specific terminology 
``non-contextual probability'' (NCP), and spells out its meaning as it applies to
the Born rule:  ``The probability is non-contextual if $p_k$ does not depend on 
the measurement with other eigenvalues.''  This statement corresponds more 
closely to our concept of ``observable independence." 

We should also point out a couple of recent approaches which claim to derive
the Born rule without making the assumption of non-contextual probability.

In one such proof, Masanes et. al. (MGM) \cite{MGM} present a general and 
comprehensive account of the axioms, in which they claim to derive their versions 
of our IV and V (which include the Born rule) from their versions of our I, II, and III, 
so that the Born rule is derived without the assumption of NCP.  This is 
accomplished by taking an operational approach, in which the notions of measurement 
and outcome probability are taken as primitive elements of the theory.   Further 
assumptions are stated, also couched in operational terms, for example (i) that 
the predictions of the theory (eg., $p_k$) are independent of the observer's choice 
for partitioning of the system into sub-systems, and (ii) the possibility of state 
estimation.  So NCP is not an assumption, and yet it is a property of the theory, 
being implicit in the Born rule.  It is not clear to us how this essentially physical 
aspect of the Born rule arises from the other axioms and the stated assumptions. 


Zurek \cite{Zurekenv} attempted a new and interesting kind of proof based on
the interaction between the environment and the pointer variable of the apparatus,
while avoiding decoherence theory, which employs the Born rule.  Instead, the Born 
rule is derived from a symmetry under joint transformations of the environmental and
pointer states.  The symmetry is called ``envariance,'' for ``environmentally-induced 
invariance.''  This work prompted a general critical analysis detailing implicit 
assumptions and open questions \cite{Schloss-Fine}.  And more specifically, it was 
criticised for implicitly assuming non-contextual probability \cite{LS.12} in the proof.  
Hence it seems that the environment-pointer mechanism does not avoid this (NCP)
assumption after all, which removes the motivation for introducing this more
complicated mechanism in the first place. 
It is notable to us that, while decoherence theory suggests that the environment
is crucial in describing the collapse {\it phenomenon}, we find no reason to 
believe that it plays a role in the Born rule itself, since the Born rule is derived 
from simpler considerations.  We believe that the Born rule is more fundamental.   


\section{Conclusions}

We have presented a derivation of the quadratic dependence of the Born rule
from the ``dynamical axioms'' (I - III), plus the single additional assumption that
the probability is ``non-contextual,'' or perhaps less ambiguously, ``observable 
independent.'' This derivation is one of many alternatives to GleasonÕs original
proof \cite{Gleason.57}.  It is aimed at maximizing the physical content and 
minimizing the formalism.  It is in the language, and at the level of a first-year
graduate text in quantum mechanics.

Strictly speaking, the assumption of observable independence is part of the
Born rule itself, as this assumption is implicit in Eq. \ref{Born rule}.
In this light, we have simply derived one aspect of the Born rule from 
another.   In principle, one could replace the Born rule as stated in axiom 
IV (through Eq \ref{Born rule}) by the alternative statement that probability is 
observable-independent.  However, one equation (\ref{Born rule}) is worth 
a thousand words, and although in some sense redundant, its elegance 
argues for maintaining the statement as is - but to be understood as part 
axiom and part derived.

\medskip





\clearpage
\begin{widetext}
\begin{appendix}

\end{appendix}
\clearpage
\end{widetext}

\end{document}